\documentclass[preprint, showpacs,preprintnumbers,amsmath,amssymb,showkeys]{revtex4}
\usepackage{amsmath}
\usepackage{bm}
\usepackage{graphicx}

\newcommand{\be}{\begin{equation}}
\newcommand{\ee}{\end{equation}}
\newcommand{\bea}{\begin{eqnarray}}
\newcommand{\eea}{\end{eqnarray}}
\newcommand{\bc}{\begin{center}}
\newcommand{\ec}{\end{center}}

\begin{document}

\title{Chemical potential as a source of stability for gravitating Skyrmions}
\author{M. Loewe}
\email{mloewe@fis.puc.cl} \affiliation{Facultad de F\'\i sica,
Pontificia Universidad Cat\'olica de Chile,\\ Casilla 306,
Santiago 22, Chile.}
\author{S. Mendizabal}
\email{smendiza@fis.puc.cl} \affiliation{Facultad de F\'\i sica,
Pontificia Universidad Cat\'olica de Chile,\\ Casilla 306,
Santiago 22, Chile.}
\author{J.C. Rojas}
\email{jurojas@ucn.cl} \affiliation{Departamento de F\'{\i}sica,
Universidad Cat\'{o}lica del Norte, Angamos 0610, Antofagasta,
Chile}

\begin{abstract}
A discussion of the stability of self gravitating Skyrmions, with a
large winding number N, in a Schwarzschild type of metric, is
presented for the case where an isospin chemical potential is
introduced. It turns out that the chemical potential stabilizes the
behavior of the Skyrmion discussed previously in the literature.
This analysis is carried on in the framework of a variational
approach using different ansaetze for the radial profile of the
Skyrmion. We found a divergent behavior for the size of the
Skyrmion, associated to a certain critical value $\mu_c$ of the
chemical potential. At this point, the mass of the Skyrmion
vanishes. $\mu_c$ is essentialy independent of gravitating effects.
The stability of a large N skyrmion against decays into single
particles is also discussed.
\end{abstract}

\maketitle

The stability of a Skyrme topological soliton coupled to gravity has
been analyzed many times in the literature
\cite{kodama,bizon,dross}. The purpose of this letter is to extend
this discussion to the scenario where Isospin chemical potential is
taken into account.

Gravitating Skyrmions turn out to have a very rich structure, with a
nontrivial spectrum for the asymptotic states \cite{bizon}, and
provide a counter example to  the ``non hair theorem" in the
construction of black hole solutions \cite{dross}. It is therefore
interesting and natural to consider such objects now in the presence
of chemical potentials associated to conserved charges.

Recently, we have discussed Skyrmions as a model for baryons in the
presence of isospin chemical potential ($\mu$), showing the
existence of a critical value $\mu_c$ where the mass of the Skyrmion
vanishes, signalizing the occurrence of phase transitions. We found
that the radial profile of the Skyrmion becomes broader as $\mu$
approaches the critical value. The nucleon spectra was also
considered in this context \cite{lmr}.

Here we present, following the article by Glendenning {\it et al}
\cite{kodama}, a variational analysis for the stability of a self
gravitating SU(2) Skyrmion for large values of the winding number
N, showing that $\mu$ plays an stabilizing role. In particular,
the parameter associated to the radius of the Skyrmion profile
starts to grow, whereas the mass diminishes as function of $\mu$.
For a certain value $\mu_c$, the radius diverges and the mass
vanishes in the same way as it happens in the non gravitating case
\cite{lmr}. We found strong numerical evidence that the value of
$\mu_c$ is independent of the parameters associated to gravity.

It was found in \cite{kodama}, that the Skyrmion mass behaves like
$O(N^2)$. Therefore gravity becomes important in the large N limit.
Actually, the scale where gravity becomes relevant is given by
$Nf_{\pi}\sim M_{\rm Planck}$. The Skyrmion mass depends on the
quotient $R_S/R \approx G (N f_{\pi})^2$, where $R$ is a measure of
the Skyrmion size in the absence of gravity, whereas $R_S$ is
essentially the Schwarzschild radius, i.e. the natural length scale
associated to gravity effects. In the variational approach, there is
no solution for values of $R_S/R$ bigger than a certain limit. If
$\mu \neq 0$, this limit becomes bigger, signalizing the stabilizing
role of the chemical potential. If $\mu \rightarrow\mu_c$, we have
no bound for $N$.

The authors of \cite{kodama} found also that their solution actually
does not satisfy the condition $M(N)\leq N M(1)$, indicating that
the solution with high winding number should decay into many
Skyrmion states with lower N. In our case, the chemical potential
also provides a source of stability near $\mu_c$.

We would like to remark that in our previous analysis \cite{lmr},
the variational approach was not appropriate due to the lack of
stability for large values of $\mu$. Therefore, a full numerical
treatment was unavoidable. However, in the case of the self
gravitating Skyrmion, a variational approach is possible for any
value of $\mu < \mu_c$. The idea is to consider a certain given
radial profile, which depends on a free parameter, minimizing then
the mass of the Skyrmion.

Let start our analysis with the nonlinear sigma-type Skyrme
lagrangian

\bea \label{lagra}L_m &=& \frac{ {\rm f}_{\pi}^2}{4} Tr\left[
D_{\mu}U D^{\mu}U^{\dagger} \right] \nonumber \\ &&+
\frac{1}{32e^2}Tr \left[ (D_{\mu}U)U^{\dagger},
(D_{\nu}U)U^{\dagger} \right]^2, \nonumber \\ &\equiv& L_2+L_4,
\label{Lag}\eea

\noindent where

\be D_{\mu}=\partial_{\mu}-{\rm i}\frac{\mu}{2}
\left[\sigma_3,U\right]\delta_{\mu,0}, \ee

\noindent is the covariant derivative that introduces the chemical
potential \cite{actor,weldon}, and

\bea U=\sigma_0\cos F(r)+i\vec{\sigma}\cdot\hat{n}\sin F(r), \eea

\noindent is the standard Hedgehog ansatz, $\vec{\sigma}$ are the
Pauli matrices, $\sigma_0$ is the identity and $F(r)$ is the radial
profile. The gravitating version (Einstein-Skyrme lagrangian) can be
obtained through the action

\be S=\int \left( L_g+L_m \right)\sqrt{-g}\; d^4x,\ee

\noindent with $L_m$ given by (\ref{lagra}) and

\be L_g = -\frac{R}{16 \pi G}, \ee

\noindent where $R$ is the Ricci scalar curvature and G the Newton
constant.

%\bea
%\vec{\sigma}\cdot\hat{n}=\sigma1\sin{\theta}\cos\phi+\sigma2\sin
%\theta \sin\phi+\sigma3\cos(\theta). \eea

The explicit form of $L_2$ and $L_4$, without $\mu$ is

\bea L_{2,0}&=&\frac{1}{2}f_{\pi}^2 \Big(g^{11}F'(r)^2 +g^{22}\sin^2
F(r)\nonumber
\\
&& +g^{33}\sin^2 F(r)\sin^2 \theta \Big), \eea

\bea L_{4,0} &=& -\frac{1}{2e^2}\Big(g^{11}g^{22}F'(r)^2\sin^2 F(r)
\nonumber
\\
&&+g^{11}g^{33}F'(r)^2\sin^2 F(r)\sin^2 \theta \nonumber
\\&& +g^{22}g^{33}\sin^4
F(r)\sin^2 \theta\Big). \eea

When we turn on the chemical potential,  $\mu$-dependent terms
appear, so that, in an obvious notation,

\be L_m=L_{2,0}+L_{4,0}+L_{2,\mu}+L_{4,\mu}. \ee

\noindent with

\bea L_{2,\mu}=\frac{1}{2}f_{\pi}^2\mu^2\sin^2 F(r)\sin^2 \theta,
\eea

\noindent and

\bea L_{4,\mu} &=&-\frac{1}{2e}\mu^2\sin^2 F(r)\sin^2 \theta
\Big(g^{11}F'(r)^2 \nonumber \\ &&+g^{22}\sin^2 F(r) \Big). \eea

The Schwarzschild metric depends on two arbitrary functions $\nu(r)$
and $\lambda(r)$, in such a way that

\bea g^{00}&=&e^{-2\nu(r)}, \;\;g^{11} =-e^{-2\lambda(r)},\nonumber\\
g^{22}&=&-\frac{1}{r^2}, \; \;  g^{33} = -\frac{1}{r^2\sin^2
\theta}. \eea

Our starting point is the Einstein equation

\be G^{\mu \nu} = -8\pi G T^{\mu\nu}.  \label{ein-sky}\ee

The matter energy-momentum tensor is given by

\be T^{\mu \nu} = -g^{\mu \nu} L_m+2 \frac{\partial L_m}{\partial
g_{\mu \nu}}.\ee

\noindent For our variational approach, where different ansaetze for
the radial profile $F(r)$ will be used, only the $T^{00}$ component
is needed

 \bea T^{00}&=& \frac{{{\it f_{\pi}}}^{2}}{2} \left(  {\frac { 2 \sin^2 F(r)}{{r}^{2}}}+{\frac {
  F'^{2}(r)}{ {e^{2\lambda
 \left( r \right) }} }} \right) \nonumber \\ && + \frac{1}{2 {e}^{2}} \left({\frac { \sin^4 F(r)}{{r}^{4}}}+{\frac {
 \sin^{2} F(r)  F'^{2}(r)}{
{e^{ 2\lambda \left( r \right) }}  {r}^{2}}} \right) \nonumber \\
&& -\frac{{\it f_{\pi}}^{2} {\mu}^{2}}{2} \sin^{2} F(r)
 \sin^2 \theta \nonumber \\ && -\frac{{\mu}^{2
}}{2{e}^{2}} \left({\frac { \sin^{2} F(r)  \sin^2 \theta F'^{2}(r)}{
{e^{2\lambda \left( r
 \right) }}  }} \right. \nonumber \\ &&  \left. +{\frac {  \sin^{4}  F
 \left( r \right)    \sin^{2}  \theta
 }{{r}^{2}}} \right).
\eea

For our purposes, we will only consider the angular average over
the unit sphere. Since we are following a variational approach,
the reasons that justify the self gravitating skyrmion stability
will be not affected by taking this angular average. In this way,
we replace $\sin^2(\theta)\rightarrow 2/3$, obtaining an averaged
energy-momentum tensor component $ \langle T^{00} \rangle$

The first component of (\ref{ein-sky}) reads

\be r^2G_0^0\equiv e^{-2\lambda}(1-2r\lambda')-1=8\pi G r^2
\langle T_0^0\rangle,\label{eins1} \ee

\noindent where the right hand term, written in terms of the
dimensionless variable  $x=e f_{\pi} r$, becomes

\bea 8\pi G r^2 \langle T_0^0\rangle &=& 4\pi G
f_{\pi}^2\left[\sin^{2}F(x)+\frac{\sin^4 F(x)}{2x^2} \right. \nonumber\\
&& -\frac{\mu^2}{2e^2 f_{\pi}^2}\sin^{4} F(x)\frac{2}{3}
 \nonumber \\ &&-\frac{\mu^2}{2e^2f_{\pi}^2}x^2\sin^{2} F(x) \frac{2}{3}
 \nonumber\\ &&+ \left(1-\frac{2\nu(x)}{x}\right)
\left(\frac{x^2}{2} +\sin^2 F(x)
\right.  \nonumber\\
&-& \left. \left. \frac{\mu^2}{2e^2f_{\pi}^2} x^2
\sin^{2}F(x)\frac{2}{3}  \right) F'^2 \right]. \eea

Following \cite{kodama}, equation (\ref{eins1}) can be written as

\be -\frac{d}{dr}[r(1-e^{-2\lambda(r)})]=-8\pi Gr^2 \langle
T_0^0\rangle.\ee

A formal integration of the previous equation leads us to the
identification

\be r(1-e^{-2\lambda(r)}) = 8\pi G \int_0^{r} \langle T_0^0\rangle
r'^2 dr'. \ee

\noindent It is natural to define the mass inside the radius r as

\be M(r)=4 \pi \int_0^{r} \langle T_0^0\rangle r'^2 dr'. \ee

\noindent Then, we can identify the left term in  (\ref{eins1}) as
$r(1-e^{-2\lambda(r)}) \Rightarrow 2G M(r)$.

\noindent In order to compare with \cite{kodama}, we introduce $R
\equiv \sqrt{3}/ef_{\pi}$ as a relevant length scale parameter. In
terms of R, we define the dimensionless quantities

\be \frac{R_S}{R}\equiv\frac{8\pi^3}{3}G(Nf_{\pi})^2,\;\;\;\;
\tilde{\mu} \equiv \mu R= \frac{\sqrt{3}\mu}{ef_{\pi}}.\ee

In what follows we will use the dimensionless variable
$\tilde{M}(x)\equiv e f_{\pi}G M(x)$. In this way, we find

\be \frac{d\tilde{M}(x)}{dx}=\left[1-\frac{2\tilde{M}(x)}{x}
\right]p(x)+q(x),\ee

\noindent where we have introduced the functions $p(x)$ and $q(x)$
according to

\bea p(x)&=&\frac{1}{4}\,\frac{{\it R_S}}{R}\, \left( {\frac
{F'(x)}{{\pi }{N}}}  \right) ^{ 2} \large( 6 \sin^2F(x)+ 3 x^{2}
\nonumber \\ && -\frac{2 \tilde{\mu}^2\,{ {x}^{2}} \sin^2F(x)}{3}
\large),
 \eea

\noindent and

\bea q(x)&=& \left( \frac{R_S}{R}\right) \frac{\sin^2 F(x)}{4 \pi^2
N^2} \left(6 x^2+3\sin^2 F(x) \right. \nonumber \\
&& -\frac{2 \tilde{\mu}^2x^4}{3}- \frac{2 \tilde{\mu}^2x^2
\sin^2F(x)}{3} \large) \eea

\bigskip

Different ansaetze may be used for the radial profile, in order to
proceed with the variational approach, fulfilling the boundary
conditions $F(0)=N\pi$  and $F(x\rightarrow \infty)=0$.
Nevertheless, the results seem to be more or less independent of the
details of the parametrization employed. A simple and natural
profile emerges from the discussion of instanton  holonomies
\cite{Manton}

\be F(x)=N\pi \left[1-\frac{x}{\sqrt{x^2+X^2}} \right]. \ee

\noindent This profile is also valid for curved spaces
\cite{Sutcliffe}. Actually the validity of this ansatz has been
proved in \cite{nitta}. Other approaches, as for example,
exponential decaying profiles $F(x)=N\pi \exp(-x/X)$ have been
used. The variational procedure leads us to the optimal value of
the free parameter $X$ in each case.

For our estimates, we will employ the first ansatz. The main reason
for this, is a better numerical convergence of the integrals which
give us the mass of the Skyrmion. We would like to remark that this
is possible when gravity effects are taken into account. In our
previous works on the stability of hadrons described as Skyrmions
and their excitations, in the presence of $\mu$, this
parametrization was only valid for small values of $\mu$. In the
present case, we do not have such instability problems when going to
higher values of $\mu$. As we claimed in the introduction, gravity
compensates the instability of the non gravitating Skyrmion.

In the large N limit, due to the rapid oscilating factor, we may
approximate $\sin^2F(x)\approx 1/2$. On the other side, since
$q(x) \propto O(1/N^2)$ and $p(x)\propto O(1)$, we can neglect
$q(x)$, getting

\be \frac{d\tilde{M}(x)}{dx}=\left[1-\frac{2\tilde{M}(x)}{x} \right]
\bar{p}, \ee

\be \bar{p}=\frac{3}{4}\,\frac{{\it R_S}}{R}\, \left( {\frac
{\dot{F}(x)}{{\pi }{N}}}  \right) ^{ 2} \left( 1+
x^{2}-\frac{\tilde{\mu}^2}{9}\,{ {x}^ {2}} \right)
 \ee

Gledenning et al \cite{kodama} found stable solutions for the
Skyrmion only for values of $R_S/R \lesssim 0.26$. This means that
for bigger values of $R_S/R$, the mass of the Skyrmion does not have
a minimum with respect to X.

\begin{figure}[h]
\includegraphics[angle=0,width=0.5\textwidth]{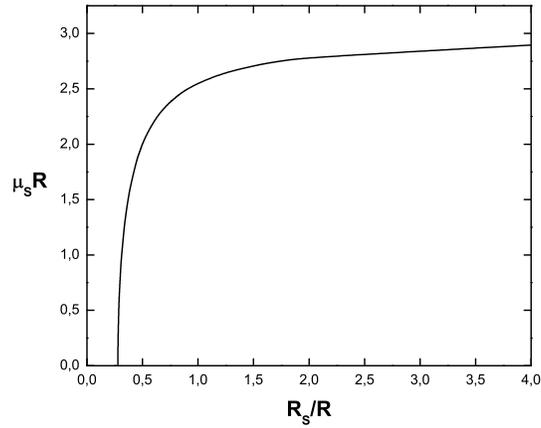}
\caption{\label{mustable} The behavior of $\tilde{\mu}_S=\mu_SR$
is shown as function of $R_s/R$. It goes asymptotically to the
critical value $\tilde{\mu}_s \rightarrow \tilde{\mu}_c$.}
\end{figure}

\begin{figure}[h]
\includegraphics[angle=0,width=0.5\textwidth]{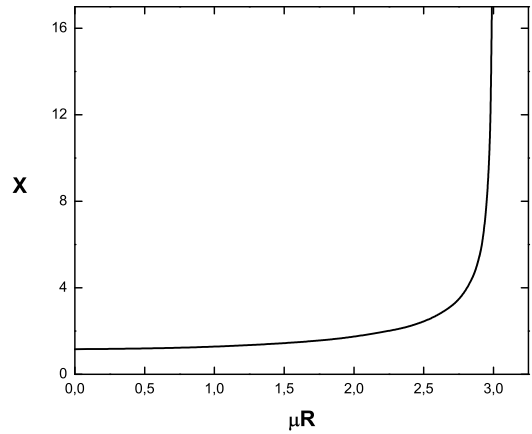}
\caption{\label{RS_R=0168} This figure shows the evolution of the
variational parameter $X$ that minimizes the mass as function of
the chemical potential. It diverges when $\tilde{\mu} \rightarrow
\tilde{\mu}_c$.}
\end{figure}

In our case, for each value of $R_S/R$, we are able to find a
stabilizing value of $\mu$ ($\mu_S$), such that the mass of the
Skyrmion has a minimum as function of X. The behavior of $\mu_S$ as
function of $R_S/R$ is shown in figure (\ref{mustable}). Notice that
$\mu_S$ asymptotically goes to a certain critical value
$\mu_c\approx0.3/R$ ($80$ MeV), where the size of the Skyrmion
diverges. This is shown in figure (\ref{RS_R=0168}), where the
divergent behavior of the parameter X is plotted as function of
$\mu$, signalizing also a divergent behavior for the radial
extension of the profile, since bigger values of X imply a broader
radial profile.

\begin{figure}[h]
\includegraphics[angle=0,width=0.5\textwidth]{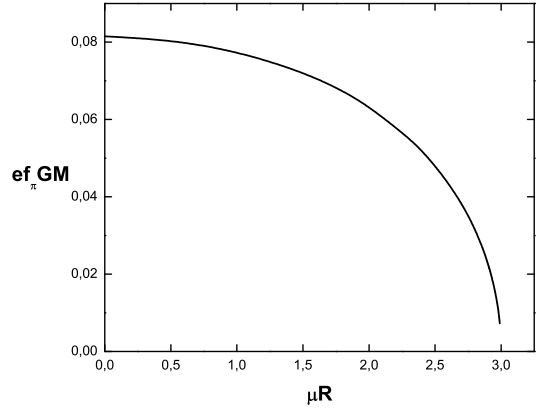}
\caption{\label{mRS_R=0168} Here we show the Skyrmion mass
evolution as function of $\tilde{\mu}$. The mass vanishes when
$\tilde{\mu} \rightarrow \tilde{\mu}_c$.}
\end{figure}

In the same way, simultaneously with these effects, it turns out
that $\tilde{M}=ef_{\pi}M$ diminishes as function of $\mu$,
vanishing at the critical value $\mu_c$. This is shown in figure
(\ref{mRS_R=0168}). It is interesting to remark that the same kind
of behavior for the mass of the Skyrmion was found in the non
gravitating scenario discussed in our previous work. This means that
the phase transition induced by the chemical potential is triggered
by the strong dynamics.

Finally, we would like to address the problem of stability of the
large N Skyrmion into particle emission. The condition for that is
$M(N)<NM(1)$. This means that

\be \frac{d\tilde{M}(N)}{d(R_S/R)}< \frac{{\rm A}}{N}.
\label{condition}\ee

\begin{figure}[h]
\includegraphics[angle=0,width=0.5\textwidth]{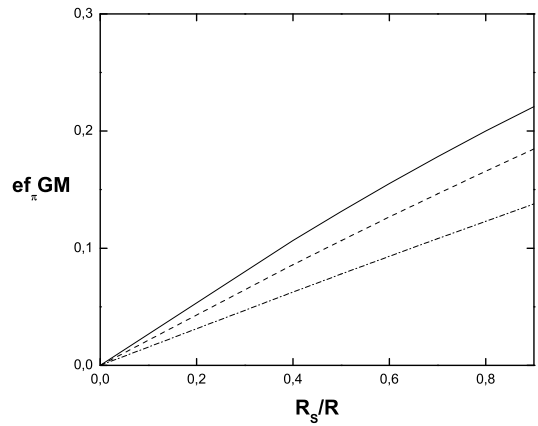}
\caption{\label{dimlessRSR} Here we show the Skyrmion mass
dependence as function of $R_S/R$ for three different values of
$\tilde{\mu}$: continuous line $\tilde{\mu}=2.5$; dashed line
$\tilde{\mu}=2.7$; dash-dotted line $\tilde{\mu}=2.85$.}
\end{figure}

\noindent From figure (\ref{dimlessRSR}) we see that the slope of
$\tilde{M}$ as function of $R_S/R$ diminishes when $\mu$ increases.
For $\tilde{\mu} \rightarrow \tilde{\mu}_c$, the slope vanishes.
This means that for any value of $R_S/R$ we can always find a value
for $\mu$, such that the condition (\ref{condition}) is fulfilled,
implying the stability against decay.

In this letter we have discussed stability conditions for self
gravitating large N $SU(2)$ skyrmions in the presence of isospin
chemical potential. Our main results show that this chemical
potential plays an important role stabilizing the solution,
extending the range of validity for the solutions as function of
$R_S/R$. We showed that the skyrmion mass vanishes at the same
critical chemical potential where the size of the Skyrmion
diverges. Finally, this scenario allows the existence of stable
large N Skyrmions against decays into single Skyrmions.

\newpage

\section*{ACKNOWLEDGMENTS}

The authors would like to thank financial support from
 FONDECYT under grants 1051067 and 1060653.

%\section*{ACKNOWLEDGMENTS}

\end{document}